# Probing electron-electron interaction along with superconducting fluctuations in disordered TiN thin films

Sachin Yadav, [1,2] Vinay Kaushik, [3] M.P. Saravanan, [3] and Sangeeta Sahoo*[1, 2]

[1] CSIR-National Physical Laboratory, Dr. K. S. Krishnan Marg, New Delhi-110012, India

[2] Academy of Scientific and Innovative Research (AcSIR), Ghaziabad- 201002, India

[3] Low Temperature Laboratory, UGC-DAE Consortium for Scientific Research, University Campus, Khandwa Road, Indore 452001, India

*Correspondences should be addressed to S. S. (Email: sahoos@nplindia.org)

In 2D disordered superconductors prior to superconducting transition, the appearance of a resistance peak in the temperature dependent resistance [$R(T)$] measurements indicates the presence of weak localization (WL) & electron-electron interaction (EEI) in diffusion channel and superconducting fluctuations in the Cooper channel. Here, we demonstrate an interplay between superconducting fluctuations and electron-electron interaction by low temperature magnetotransport measurements for a set of 2D disordered TiN thin films. While cooling down the sample, a characteristic temperature $T^*$ is obtained from the $R(T)$ at which superconducting fluctuations start to appear. The upturn in R(T) above T* corresponds to WL and/or EEI. By the temperature and field dependences of the observed resistance, we show that the upturn in $R(T)$ originates mainly from EEI with a negligible contribution from WL. Further, we have used the modified Larkin's electron-electron attraction strength $\beta(T/T_c)$, containing a field induced pair breaking parameter, in the Maki-Thompson (MT) superconducting fluctuation term. Here, the temperature dependence of the $\beta(T/T_c)$ obtained from the magnetoresistance analysis shows a diverging behavior close to $T_c$ and it remains almost constant at higher temperature within the limit of $ln(T/Tc) <1$. Interestingly, the variation of $\beta(T/T_c)$ on the reduced temperature ($T/T_c$) offers a common trend which has been closely followed by all the concerned samples presented in this study. Finally, the temperature dependence of inverse phase scattering time ($\tau_\phi^{-1}$), as obtained from the magnetoresistance analysis, clearly shows two different regimes; the first one close to $T_c$ follows the Ginzburg-Landau relaxation rate ($\tau_{GL}^{-1}$), whereas, the second one at high temperature varies almost linearly with temperature indicating the dominance of inelastic electron-electron scattering for the dephasing mechanism. These two regimes are followed in a generic way by all the samples in spite of being grown under different growth conditions.

# I. INTRODUCTION

In a superconductor, the transition from the metallic state occurs in two phases: first, the order parameter is established with a finite amplitude at the critical temperature ($T_c$), and then, the formation of the global phase coherent state at the characteristic temperature $T_{BKT}$ [1-3]. In 2D disordered superconductors above the $T_c$, electrical transport properties are mainly controlled by the quantum corrections to the conductivity (QCC) [4, 5]. The QCC can broadly be summarized into two parts; first is the weak localization (WL) due to quantum interference of complementary electron waves travelling in a closed loop but opposite in direction, second is the disorder induced electron-electron interaction (EEI) [6-8]. Further, quantum corrections originating from EEI can be divided into two parts: The first part includes Coulomb interaction between the particles with close momenta in the diffusion channel (ID), and the second part includes Coulomb interaction between the particles having opposite momenta in the Cooper channel. Correction to the conductivity due to the Cooper channel becomes important, once the system transits to superconducting state. However, corrections to the conductivity arising from the channel of the Cooper pairs are further divided into three superconducting fluctuations, namely, Aslamazov–Larkin (AL) contribution which is mainly due to the participation of Cooper pairs in conduction through parallel channel [9], Maki-Thompson (MT) contribution which reflects the influence of superconducting fluctuations on normal quasiparticles [10, 11] & Density of states (DOS) contribution which originates due to the formation of superconducting pairs that lead to the reduction in the density of states for normal electrons [12]. Here, superconducting fluctuations related to AL & MT give positive and DOS gives negative contribution to the conductivity under zero magnetic field [13]. Whereas, WL+ID together offers a negative contribution to the conductivity. However, superconducting fluctuations & WL are very much sensitive to the magnetic field and they get suppressed under the application of magnetic field, whereas the contribution from the EEI remains unaffected under a high magnetic field [14].

In this article, we have studied an interplay between superconducting fluctuations in the Cooper channel and electron-electron interaction in the electronic diffusion channel by low temperature magnetotransport measurements for a set of disordered TiN thin film samples that are in 2-dimensional limit. A detailed study using all the aforesaid quantum corrections to the conductivity is reported by Baturina *et al* [13] for disordered superconducting TiN thin films of thickness in the same range (<5 nm) as that of the samples presented in this work. The observations and conclusions on the zero field *R(T)* measurements for the present work closely follow the reported results from Ref. 13. However, the study presented here extends up to an advanced level where an external magnetic field, applied perpendicular to the sample plane, is used to probe only the EEI by suppressing the other relevant mechanism like WL which may lead to an upturn in zero-field *R(T)* measurements. By the analysis of magnetoresistance results, we show here that indeed electron-electron interaction is dominant for these superconducting films above the transition temperature.

Here, samples are produced by using previously demonstrated substrate mediated nitridation technique where the annealing temperature and the film thickness were varied. The set of samples selected for this study includes samples grown with different annealing temperature and also samples varying in film thickness. However, the results from the transport measurements carried out on these varieties of samples follow similar characteristics when we consider the temperature dependent resistance measurements [*R(T)*] and/or the magnetoresistnace (*MR*) measurements. For example, while cooling down from room temperature, zero-field *R(T)* characteristics for all the samples feature a resistance dip at a specific temperature $T_{min}$ which is then followed by an upturn with negative *dR/dT* slope and finally reaching to a resistance peak at temperature $T_{max}$ and further cooling leads to superconductivity related drop in resistance. All these distinct regions, characterized mainly by the sign of the slope *dR/dT*, are present in zero-field *R(T)* for each of the samples presented here. Further, we have obtained the characteristic temperature *T\** at which superconducting fluctuations start to appear and the experimental *R(T)* starts to deviate from the WL+ID path. The characteristic temperatures *T\**, $T_{max}$ & $T_{min}$ and the resistance peak have been explicitly monitored with respect to magnetic field and we find that the contribution from WL to the QCC is very weak compared to that from EEI and hence, EEI can be considered as the main mechanism behind the upturn and the resistance peak observed in the *R(T)*. Further, the MR measurements show positive MR at temperature far above $T_c$ and no trace of negative MR is observed even above *T\** where superconducting fluctuations can be ignored. As negative MR is a hallmark for WL [15], MR measurements too indicate that the contribution from WL is not significant.

As far as the superconducting fluctuations are concerned, above $T_c$, the MT correction is the most dominant contribution to QCC. The strength of the MT contribution is generally expressed by the electron-electron attraction strength *β(T/T_c)* which was originally proposed by Larkin [16]. Here, *β(T/T_c)* varies differently for (*ln(T/T_c)<< 1*) and (*ln(T/T_c) >>1*). However, it is problematic to evaluate the value of *β(T/T_c)* in the intermediate temperature regime and in this regime, no clear report/guidance about the form of the *β(T/T_c)* is available in the literature. However, a modified magnetic field dependent *β(T/T_c)* has been proposed by Lopes dos Santos and Abrahams in the literature which is valid for low temperature regime [17] . In this study, we have considered the modified *β(T/T_c)* which depends on a pair breaking parameter '*δ*' via the phase scattering time ($\tau_\phi$) which is obtained from the MR analysis . The dependence of *β(T/T_c)* on the reduced temperature (*T/T_c*) shows a diverging behavior close to $T_c$, whereas, it behaves almost independent of temperature for the regime which is a little far from $T_c$ but still satisfying the condition (ln(*T/T_c*)<1). Interestingly, the variation of *β(T/T_c)* on *T/T_c* offers a common path which is being followed by all the samples in a collective/universal manner. Furthermore, the inverse phase relaxation time ($\tau_\phi^{-1}$) as obtained from the MR varies generically on reduced temperature (*T/T_c*) for all the samples. Close to $T_c$, superconducting fluctuations dominate and $\tau_\phi^{-1}$ follows Ginzburg-Landau relaxation rate ($\tau_{GL}^{-1}$) and at higher temperature, the phase relaxation time varies almost linearly with temperature indicating the dominance of inelastic electron-electron scattering for the dephasing mechanism.

## II. EXPERIMENTAL

We have employed undoped Si (100) substrate covered with 80 nm $Si_3N_4$ dielectric spacer layer grown by low pressure chemical vapor deposition (LPCVD). Initially, substrates went through the standard cleaning process involving sonication in acetone & isopropanol bath for 15 minutes each. Thereafter, cleaned substrates were loaded into the Ultra high vacuum (UHV) chamber for pre heating at about 820°C for 30 minutes to remove adsorbed or trapped organic molecules on the surface of the substrate. The cleaned substrates were then transferred *in situ* to the sputtering chamber where a thin layer of Ti was deposited on the substrate by using dc magnetron sputtering of Ti target (99.999% purity) in the presence of high purity Ar (99.9999%) gas. Sputtering of Ti target was done with the base pressure less than $1.5 \times 10^{-7}$ Torr. Finally, Ti deposited substrates were transferred *in situ* to an UHV chamber for annealing. Ti thin film deposited substrates were annealed at different annealing temperatures of about (~820°C, ~780°C & ~750°C) for 2 hours at a pressure less than $5 \times 10^{-8}$ Torr. During the annealing process, Ti transformed into TiN by the substrate mediated nitridation technique [18-21] where $Si_3N_4$ substrate decomposed into Si (s) & N (g) atoms and due to high affinity of titanium towards the both, formation of superconducting TiN as the majority phase along with & the non-superconducting minority phase $TiSi_2$ took into place. However, more detail about substrate mediated nitridation technique has been reported elsewhere [20]. For carrying out the electrical transport measurements at low temperature, TiN thin film based multi-terminal devices were fabricated by using stainless steel shadow mask to pattern the TiN thin films based superconducting channel. We have used a complimentary separate shadow mask to make the contact leads for voltage and currents probes. The contact leads were made of Au (80-100 nm)/Ti (5 nm) deposited by dc magnetron sputtering. Low temperature resistivity measurements were carried out using a 16 T/2 K Physical Properties Measurement System (PPMS) of Quantum Design, USA at UGC-DAE CSR Indore with an excitation of 100 nA. The morphological characterizations of the samples were performed by using atomic force microscope (AFM) from BRUKER, Dimension ICON, in tapping mode from IIT, Mandi, India.

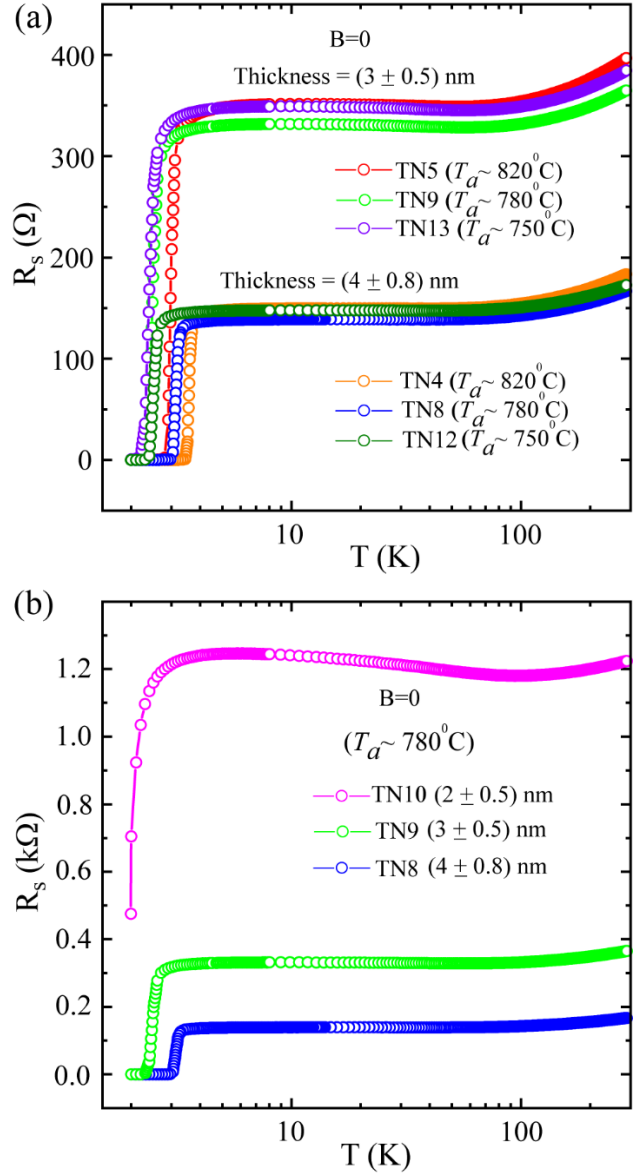

***Fig.1:*** *Semi logarithmic presentation of temperature dependent resistance [R(T)] characteristics for TiN thin films having different annealing temperature ($T_a$) and film thickness. Here, resistance $R_s$ is in Ohm/Square. (a) Two sets of zero field R(T) characteristics measured from room temperature down to 2K with each set containing three different samples of same thickness but with different annealing temperature (b) Transport properties of three samples grown at same annealing temperature but different film thicknesses.*

## III. RESULTS AND DISCUSSION

Temperature-dependent resistance $R(T)$ measurements have been carried out on TiN thin films of dimensions 1100 μm (length) × 500 μm (width) by using the conventional four-probe geometry. Further, to investigate the role of annealing temperature ($T_a$) and film thickness on the transport properties, we have fabricated TiN thin films by keeping one growth parameter fixed and altering the other. In Fig. 1(a), we have presented the zero field $R(T)$ characteristics of six different samples from room temperature down to 2 K in a semi-logarithmic scale. Based on the influence of the film thickness on the electronic properties, $R(T)$s of these six TiN thin film samples are divided into two sets, where one set of $R(T)$ belongs to the film thickness of about 3 nm and the other set corresponds to 4 nm as indicated in Fig. 1(a). We have observed that the variations in the $T_a$ from 820°C to 750°C (TN4, TN8 & TN12) while keeping the film thickness fixed at about 4 nm have very little influence on the normal state resistance ($R_N$) and corresponding $R(T)$s almost overlap with each other in their normal state. Similar behavior is observed for the film thickness of about 3 nm, where all the related $R(T)$ characteristics (TN5, TN9 & TN13) show resemblance in their $R_N$ and its variation from room temperature down to the temperature just before transiting to the superconducting (SC) state. As it is evident in Fig. 1(a), a change in the film thickness from 4 nm to 3 nm shows a significant change in the $R_N$ for any particular $T_a$ (820°C or 780°C or 750°C) considered here. Therefore, particularly in the normal state, variation in the film thickness has a greater influence on the transport properties than the influence of $T_a$. However, superconducting properties such as critical temperature ($T_c$), transition width etc. depend strongly on $T_a$ as it is apparent from Fig. 1(a) that normal metallic to superconductor transition shifts towards lower temperature with broader transition width for reducing the $T_a$ from 820°C to 780°C and further to 750°C. In order to observe the effect of thickness on the overall transport properties, we have collected a set of $R(T)$ measurements for three samples TN8, TN9 & TN10 having different film thicknesses of about 4 nm, 3 nm & 2 nm, respectively and have been grown with a fixed $T_a$ of about 780°C as shown in Fig. 1(b). With the reduction in the film thickness, $R_N$ starts to increase while $T_c$ shifts towards the lower temperature side. Here, the samples TN8

& TN9 undergo a complete superconductivity, however, with a reduction in the film thickness from 3 nm to 2 nm, sample TN10 shows partial superconductivity as observed in Fig. 1(b). Further, we have investigated the zero-field *R(T)* characteristics in more detail for the set of samples (TN8, TN9, TN10) presented in Fig. 1(b). A narrow temperature window for *R(T)* is considered for each of the samples in this set for emphasizing the close vicinity of the metal-superconductor transition and they are shown in the main panels of Fig. 2(a-c). The details about the *R(T)* variation in the normal state for all the three samples are highlighted in the insets of Fig. 2(a-c) that show different regions mainly based on the slope *dR/dT* from the measured *R(T)*. For all the samples while cooling down from room temperature, resistance decreases with decreasing temperature till it reaches to a minimum at the characteristic temperature $T_{min}$ indicating a metallic behavior with positive *dR/dT*. Further lowering the temperature, an upturn in *R(T)* appears where resistance starts to increase and reaches to a maximum at the temperature $T_{max}$. In this regime between $T_{min}$ and $T_{max}$, the negative *dR/dT* indicates an insulating or semiconducting type of behavior. At temperature below $T_{max}$, resistance drops sharply and the superconducting fluctuations take over. For example, the thinnest sample TN10 with about 2 nm thickness shows metallic behavior with positive *dR/dT* at $T > T_{min}$=95 K and an upturn accompanied by a resistance peak with negative *dR/dT* for the temperature window from 95 K to 5.8 K ($T_{max} \sim 5.8$ K) as shown in Fig. 2(a). Finally, below 5.8 K ($T_{max}$), resistance starts to drop sharply as the sample transits to superconducting state.

Similarly, while cooling from room temperature, a resistance minimum followed by an upturn and superconducting drop appear in the *R(T)* for other two samples (TN9 & TN8) as displayed in the insets of Fig. 2(b) & (c), respectively. Here, the slope of the upturn increases with decreasing film thickness. This is expected when EEI plays the dominant role, as reduction in thickness introduces more disorder and EEI increases with disorder. An upturn in resistance can originate from granularity also. However, for the granular superconducting systems, no systematic reduction in $T_c$ occurs for decreasing thickness [22] but in the present work, as shown in Fig. 1(c) and also in Table-1, we observe that the $T_c$ decreases systematically with the decreasing thickness and at the same time, the transitions remain sharp. This indicates the granularity might not be the reason behind the observed upturn in the zero-field *R(T)* [23]. Further, the surface morphology, as observed through atomic force microscopy (AFM) images shown in Fig. S7 in the Supplemental Material [24], does not clearly indicate the granular nature of the films as the images reflect the surface roughness rather than isolated grains.

Generally, WL and EEI play major roles for the appearance of resistance peak and upturn in *R(T)* for two-dimensional homogeneously disordered materials [4, 8, 25, 26]. However, the dimensionality of the system is very much sensitive to WL & EEI and superconducting fluctuations, therefore characteristic length scales such as the superconducting coherence length $\xi_{GL}(0) = \left[\frac{\phi_0}{2\pi T_c \left|\frac{dH_{c2}}{dT}\right|_{T_c}}\right]^{1/2}$ with $\phi_0$ as the flux quantum and thermal coherence length $L_T = \sqrt{2\pi\hbar D/(k_B T)}$ with $k_B$ as the Boltzmann constant and *D* as the diffusion coefficient should be less than the film thickness. Moreover, for all the samples presented here, the film thickness is less than the characteristic superconducting coherence length (~9 nm) and the thermal coherence length (~9 to ~13 nm at 100 K). Therefore, quantum corrections to the conductivity that are applicable in 2D materials can be considered here for understanding the origin of the upturn and the related resistance peak appearing in the zero-field *R(T)* characteristics. Furthermore, the sample specific characteristic parameters like $T_a$, film thickness (*d*), sheet resistance ($R_{max}$) at $T_{max}$ before superconducting transition, superconducting critical temperature ($T_c$) as obtained by QCC fit, sheet resistance at 300 K ($R_{300\,K}$), upper critical field at $T = 0\,K$ ($B_{c2}(0)$), Diffusion constant ($D$) & the thermal coherence length ($L_T$) at 100 K for all the TiN samples are listed in Table 1.

*Table 1: Specific parameters for the samples presented in the work:*

| Samples | $T_a$, (°C) ± 10°C | $d$, (nm) | $T_C$, (K), (QCC) fit | $R_{max}$, (Ω) | $R_n^{300K}$, (Ω) | $B_{c2}(0)$, (T) | $D$, ($cm^2 s^{-1}$) | $l_T$ (100 K), (nm) |
|---|---|---|---|---|---|---|---|---|
| TN4 | 820°C | 4 ± 0.8 | 3.55 | 150 | 186 | 4.41 | 0.612 | 13 |
| TN5 | 820°C | 3 ± 0.5 | 2.93 | 352 | 400 | 5.5 | 0.405 | 11 |
| TN8 | 780°C | 4 ± 0.8 | 3.1 | 139 | 168 | 4.0 | 0.589 | 13 |
| TN9 | 780°C | 3 ± 0.5 | 2.43 | 332 | 367 | 4.31 | 0.429 | 11 |
| TN10 | 780°C | 2 ± 0.5 | 1.93 | 1246 | 1227 | 6.0 | 0.245 | 9 |
| TN12 | 750°C | 4 ± 0.8 | 2.47 | 148 | 174 | 3.31 | 0.567 | 13 |
| TN13 | 750°C | 3 ± 0.5 | 2.35 | 349 | 386 | 4.41 | 0.405 | 11 |

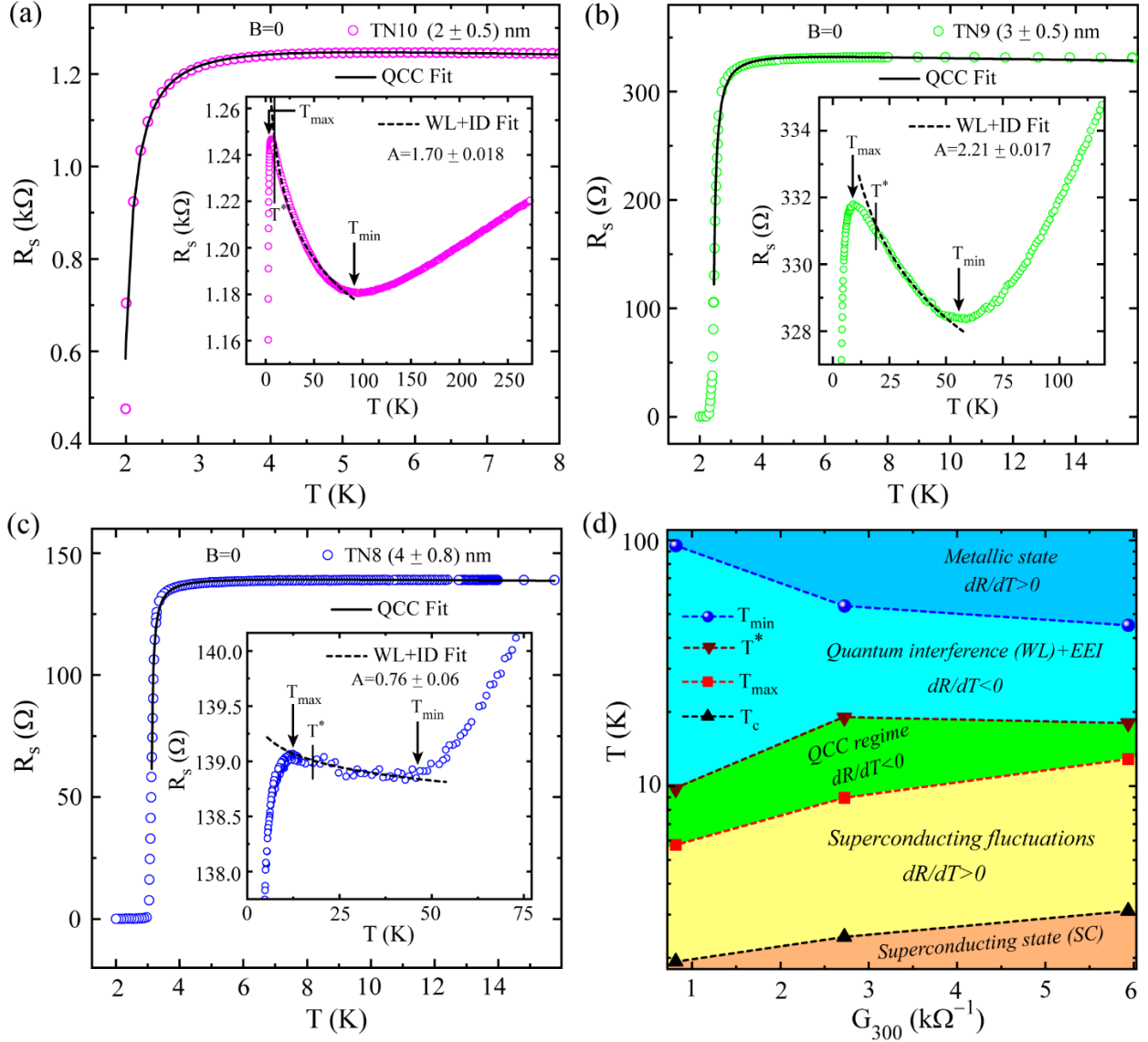

***Fig.2:*** *Investigation of zero-field R(T) data in more detail by using QCC theory for individual TiN thin films grown at same annealing temperature ($T_a$ = 780 °C) but with different thickness. (a-c) Main panels: QCC fitting to R(T) data for a selective range of temperature in the vicinity of the normal metal to superconductor transition. Insets: R(T) for a wider range of temperature demonstrating an upturn with a maximum in resistance appearing at $T_{max}$ as well as a minimum in resistance at $T_{min}$. $T_{max}$ represents the onset of resistance drop causing the superconducting transition and $T_{min}$ represents the deviation from the metallic behavior on further lowering the temperature. The region between $T_{min}$ and $T_{max}$ is fitted with (WL+ID) model as shown by the black dashed curves. (d) A Temperature-conductance phase diagram constructed by the characteristic temperatures and conductance measured at 300 K for the samples shown in (a-c). With the help of the characteristic temperatures $T_{min}$, $T_{max}$, $T_c$ & $T^*$, different regimes from metallic state to superconducting state are defined and highlighted in the phase diagram. $T_{min}$, $T_{max}$ & $T^*$ are shown in the insets of (a-c), where $T_{max}$ & $T_{min}$ correspond to the temperature points with maximum and minimum resistance, respectively, and $T^*$ is the temperature point just above the $T_{max}$ at which (WL+ID) fit starts to deviate from the experimental data. Here, $T_c$ values are extracted from the QCC fitting performed on the R(T) data for individual samples.*

Further, the presence of non-superconducting minority phases like $TiSi_2$ & elemental Si in these substrate mediated TiN thin films [19] induces disorder and disordered 2D systems are ideal candidates where WL & EEI in diffusion channel (ID) play prominent roles for the appearance of an upturn and corresponding resistive peak at $T_{max}$ in the zero-field $R(T)$. Inclined to this, we have followed here the 2D treatment with WL and EEI to address the observed upturn in the zero-field $R(T)$.

In 2D case, WL and EEI (ID) contribution to the conductivity can be obtained as [12, 26],

$$\frac{\Delta G^{WL}(T)+\Delta G^{ID}(T)}{G_{00}} = A \ln\left[\frac{k_B T \tau}{\hbar}\right] \qquad (1)$$

$$R(T) = \frac{1}{(\Delta G^{WL}(T)+\Delta G^{ID}(T)+1/R(T=10\ K)} \qquad (2)$$

with $G_{00} = e^2/(2\pi^2\hbar)$. Here, $A$ is a proportional constant and $\tau$ is the electron mean free time and they are considered as fitting parameters. From Eq. (1), corrections due to both WL and EEI to the conductivity vary logarithmically with the temperature and accordingly, the experimental $R(T)$ data for the temperature window from $T_{max}$ to $T_{min}$ is fitted by using Eq. (2). The related fits in the upturn region with negative $dR/dT$ are shown by the solid cyan curves in the insets of Fig. 2(a-c) for the aforementioned three samples TN10, TN9, TN8, respectively. The fitted curves (black dashed curves) show a reasonably good agreement with the experimental data, indicating the possibility of WL & EEI (ID) to be responsible for the observed upturn and the resistance peak in the $R(T)$. The extracted values of $A$ from the fitting for all the samples are coming less than 3 which is expected for the homogenously disordered thin films [26]. Interestingly, if we look into more detail on the WL+ID fitting from $T_{min}$ to $T_{max}$, we observe that the fit deviates from the experimental points before reaching the $T_{max}$. The temperature at the deviation point is marked as $T^*$ and the vertical dashed lines in the insets of Fig. 2(a-c) correspond to $T^*$ which occur at 9.8 K, 19 K, and 18 K for the samples TN10, TN9 & TN8, respectively. The deviation point is very clear in the inset of Fig. 2(b) for the sample TN9. Actually, the deviation relates to the onset of superconducting fluctuations that start to contribute to the conductivity corrections along with the WL and EEI. Therefore, the region between $T^*$ and $T_{max}$ shows the possibility of the co-existence of superconducting fluctuations, mainly the Maki-Thompson (MT) contributions, along with some contribution from WL & EEI. Close to $T_c$, superconducting fluctuation related to the Aslamazov-Larkin (AL) term contributes significantly in addition to the MT [9-11]. Further, the reduction in density of states (DOS) due to the Cooper pair formation contributes also to the superconducting fluctuations [12]. Therefore, the relevant contributions from superconducting fluctuations are [9-12],

$$\frac{\Delta G^{AL}(T)}{G_{00}} = \frac{\pi^2}{8} \cdot \left[\ln\left(\frac{T}{T_c}\right)\right]^{-1} \quad (3)$$

$$\frac{\Delta G^{MT}(T)}{G_{00}} = \beta(T/T_c) \cdot \ln\left[\frac{k_B T \tau_\phi}{\hbar}\right] \quad (4)$$

$$\frac{\Delta G^{DOS}(T)}{G_{00}} = \ln\left[\frac{\ln(T_c/T)}{\ln(k_B T_c \tau/\hbar)}\right] \quad (5)$$

Here, $\tau_\phi$ introduces phase breaking processes due to mainly inelastic scattering (as spin-flip scattering can be ignored) [27, 28] and $\beta(T/T_c)$ relates to the strength function characterizing electron-electron interaction which has been introduced by Larkin [16]. Summing up all the aforementioned contributions related to the total quantum corrections to the conductivity (QCC), the experimental data from $T^*$ (deviation point from WL+ID fitting) to the lowest available temperature can be fitted by using the following equation,

$$R(T) = \frac{1}{(\Delta G^{WL}(T) + \Delta G^{ID}(T) + \Delta G^{AL}(T) + \Delta G^{DOS}(T) + \Delta G^{MT}(T)) + 1/R(T=10\,K)} \quad (6)$$

The black solid curves are the fits to the experimental data by using Eq. 6 for the samples TN10, TN9 and TN8 as shown in Fig. 2(a-c), respectively. The fits follow the experimental data nicely indicating the existence of superconducting fluctuations (mainly MT) above $T_{max}$ along with the quantum interference (WL) and EEI. However, the contribution of WL+EEI can't be neglected in the region from $T^*$ to $T_{max}$, though this region is mainly dominated by superconducting fluctuations. The superconducting critical temperature ($T_c$) is obtained from the MT contribution in the QCC fitting [10, 11]. Moreover, the samples belonging to other $T_a$ (820°C & 750°C) demonstrate the same behaviour in $R(T)$ as that is observed for $T_a$

= 780°C and the corresponding $R(T)$ fittings by using Eq. 6 are shown in Fig. S1 in the Supplemental Material (SM) [24]. Finally, the characteristic temperature points such as $T_{min}$, $T_{max}$ & $T^*$ are extracted from the $R(T)$ data presented in Fig. 2(a-c) along with the $T_c$ from QCC fit and plotted them in Fig. 2(d) against sample conductance measured at 300 K. Initially, resistance follows linearly with temperature ($R \propto T$) from room temperature to $T_{min}$, where $(dR/dT)$ becomes positive and marked as a metallic state. Further, by lowering the temperature from $T_{min}$, resistance deviates from the linear behavior and starts to rise and $(dR/dT)$ becomes negative till $T_{max}$. However, this rise in resistance with the decrease in temperature is mainly dominated by WL & EEI till $T^*$ and below $T^*$, theoretical fitting of (WL+ID) starts to deviate, therefore this region from $T_{min}$ to $T^*$ is marked as a quantum interference (WL) & EEI regime in Fig. 2(d). Furthermore, the region from $T^*$ to $T_{max}$ is marked as the *QCC* regime, where combinedly superconducting fluctuations and quantum interference contribute to total quantum correction to the conductivity. However, the region from $T_{max}$ to $T_c$, where $(dR/dT)$ again becomes positive is mainly dominated by superconducting fluctuations and below $T_c$, the state is marked as the superconducting state. Summarizing, the full scale $R(T)$ from 300 K down to 2K is divided into four distinct regimes starting from the metallic state (the light blue region) to WL+EEI regime (the cyan region) to *QCC* (the green region) to superconducting fluctuations (the yellow region) to finally the superconducting state (the orange color). Further, the data in Fig. 1(b) is replotted with respect to the dimensionless conductance $G/G_{00}$ in Fig. 3 by using semi-logarithmic scale. The logarithmic temperature dependence of conductance as shown in Fig 3 confirms the two dimensionality of the samples considered for this study [8, 29] and rules out the 3D theory, where conductance $G$ varies with temperature $T$ as, $G \propto \sqrt{T}$ [30]. The logarithmic temperature dependence of conductance is the signature for the presence of WL & EEI in 2D systems [8, 29] and similar behaviour is also observed for the samples annealed at 820°C & 750°C as shown in the SM (Fig. S2) [24]. The linear fits (black dashed lines) marked in Fig. 3 show the deviation from the linearity around 10 K for the sample TN10 and around 18 K for the sample (TN9 & TN8) and these deviation points in temperature resemble with the value of *T\** shown in Fig. 2(a-c) as obtained by the fit for WL+ID contribution to the conductivity theory.

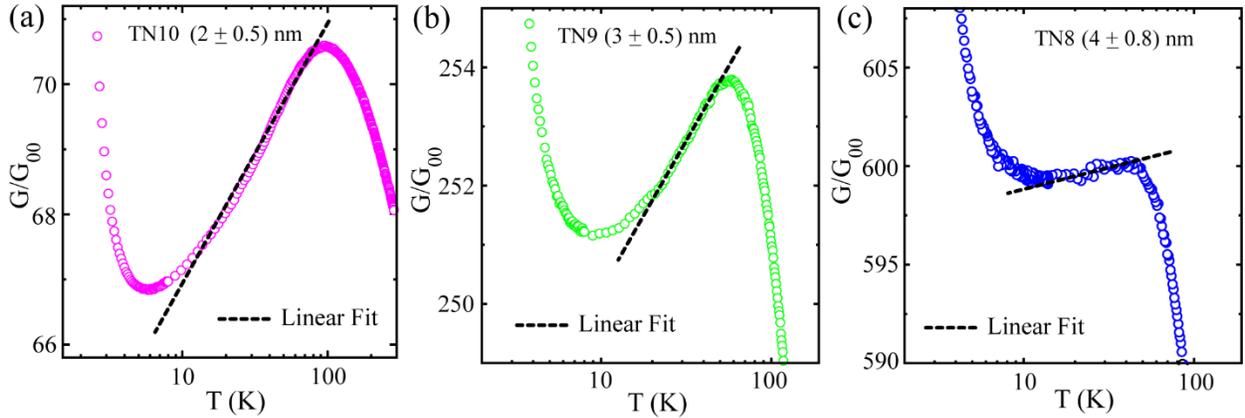

*Fig. 3: Logarithmic temperature dependence of conductance presented in semi logarithmic scale for the samples TN10, TN9 & TN8 in (a, (b) and (c), respectively. The black dashed lines are the linear fit to the experimental data.*

## *A. R(T) measurements under perpendicular magnetic field*

As quantum interference phenomenon like WL is very sensitive to the magnetic field, we have carried out the *R(T)* measurements in the presence of magnetic field applied perpendicular to the sample plane. External magnetic field can be used as a tool to distinguish the quantum phenomena WL & EEI through magnetoresistance (MR) & *R(T)* measurements. When *R(T)* measurements are carried out under perpendicular magnetic field, a relatively small field can destroy the weak localization effect but the contribution from EEI remains unaffected [14]. As already discussed, in 2D, zero-field resistance depends logarithmically with temperature for both WL and EEI. But, with the application of magnetic field, EEI is only responsible for logarithmic *R(T)* dependence as magnetic field destroys WL [14]. Here, in order to find out the actual mechanism behind the upturn observed in *R(T)* and also to find out the contribution only from EEI, we have carried out the *R(T)* measurements under external magnetic field for a sample (TN10A) selected from the same batch of TN10 and the corresponding field dependent *R(T)* is shown in Fig. 4.

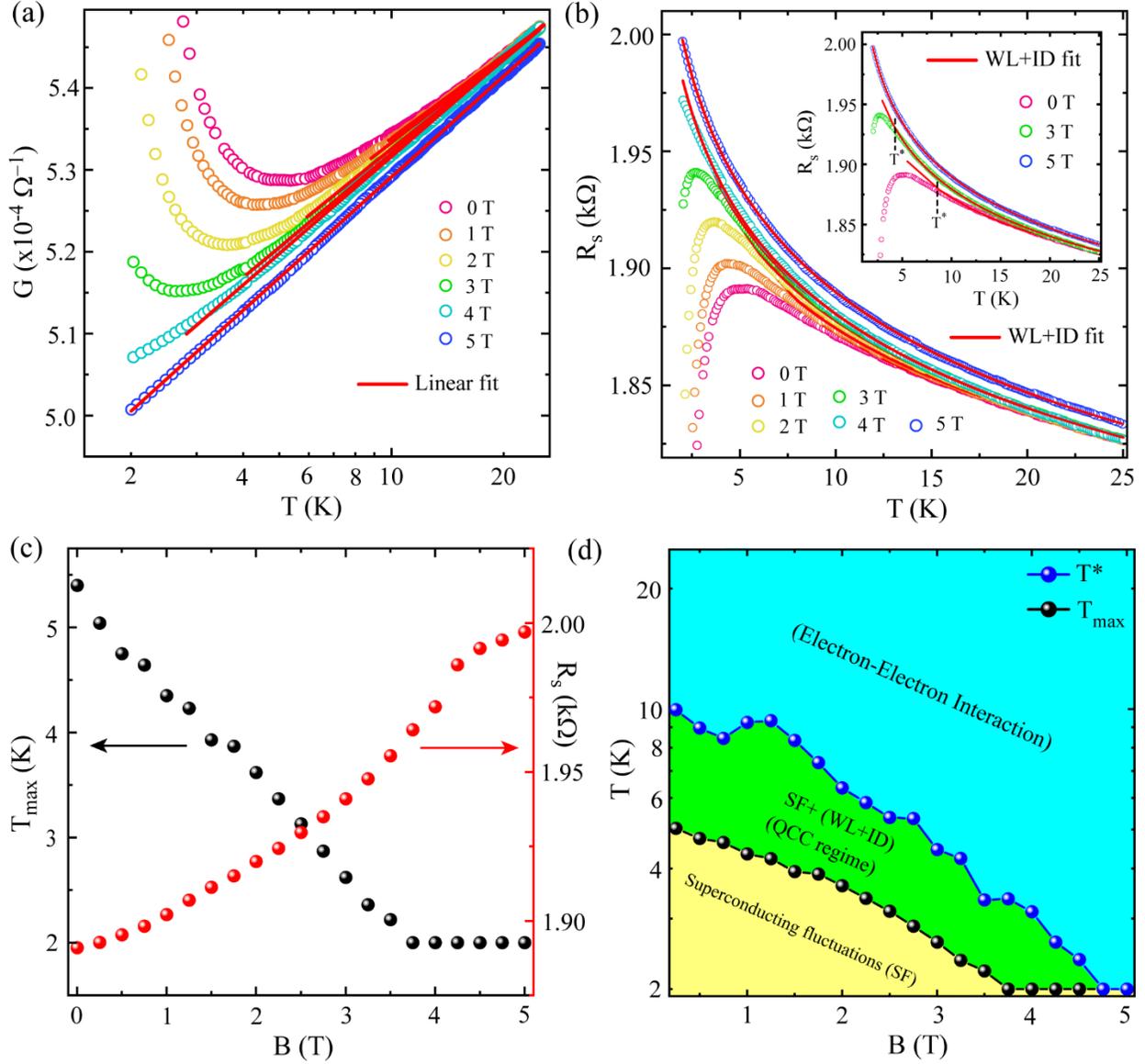

*Fig. 4:* *Field-dependent R(T) measurements for the sample TN10A (from the same batch of TN10) with magnetic field applied perpendicular to the sample plane. (a)Temperature dependence of conductance (G) in semi- logarithmic scale under various applied magnetic field. The red solid lines are the linear fits indicating the temperature dependence for the conductance remains logarithmic even after applying magnetic field. (b) Corresponding R(T)s measured under the magnetic field as mentioned in (a). The field dependent R(T)s are fitted with the contribution from (WL+ID) theory by using Eq. (2) and the fits are shown by the solid red curves. Inset: Three selective R(T)s measured under 0T, 3T & 5T from the main panel for a clearer view. For a particular field, the deviation of the (WL+ID) fit from the experimental data occurs at $T^*$ which is marked by the dashed vertical line. (c) The variation of $T_{max}$ & the corresponding maximum resistance value with the applied magnetic field. Here, the saturation in $T_{max}$ after 3.75T appears due to the temperature limitation of the measuring instrument which is limited by 2 K as the lowest achievable temperature. (d) B-T phase diagram obtained from the extracted $T^*$ & $T_{max}$ from the field-dependent R(T) data.*

First, we have considered the temperature dependence of conductance $G$ ($1/R_s$) measured under external magnetic field and the same is presented in Fig. 4(a) using a semi-logarithmic scale. The excellent agreement with the experimental data by the linear fits as presented by the red solid lines confirm the logarithmic temperature dependence for the conductance which becomes more prominent with increasing field. Due to the presence of superconducting fluctuations at relatively lower magnetic fields (1T, 2T & 3T), linear fit follows the experimental points at higher temperature region and deviates at T*. Whereas, for higher magnetic field, the linear fit follows almost the entire experimental range down to the lowest temperature (~2 K). The dominance of logarithmic temperature dependence for conductance at higher magnetic field indicates the presence of EEI. Further, to observe a clear distinction between $T_{max}$ & $T^*$ under an external magnetic field, we have presented the corresponding $R(T)$ in Fig. 4(b) by using a linear scale representation. For the applied field up to 3 T, resistance drop related to superconductivity is observed. Whereas, at 4 T and above, resistance continues to increase with temperature down to the lowest accessible temperature 2K indicating an insulating type of behavior as evident from Fig. 4(b). Here, we have fitted the resistive upturn from 25 K to the lower temperature regime by using Eq. 2 which deals with (WL+ID) correction for 2D disordered metal and the fits are represented by the red solid curves. For clarity, three representative $R(T)s$ measured at 0 T, 3 T and 5T are shown selectively in the inset of Fig. 4(b). Here, the deviation of the WL+ID fit from the experimental data is evident for 0 T and 3 T and the related temperature $T^*$ is marked by the vertical dashed lines. However, for 5 T field, the fit extends the whole experimental range down to the lowest temperature (2 K). Therefore, $T^*$ shifts towards lower temperature with increasing magnetic field. Here, the upturn region from $T^*$ to $T_{max}$ can be attributed to the presence of superconducting fluctuations along with WL & EEI contributions. Further, when $T^*$ is not distinguishable from $T_{max}$ as in the case of higher field, the magnetic field destroys superconducting fluctuations & WL contribution but EEI remains prominent and upturn in $R(T)$ gets stronger. Therefore, at high magnetic field (5T), the upturn region from $T_{min}$ to $T_{max}$ is mainly due to the presence of EEI.

Further, $T_{max}$ and its respective resistance values under the presence of an external magnetic field are displayed in Fig. 4(c). As observed in Fig. 4(b), with increasing magnetic field, $T_{max}$ shifts towards lower temperature. Above 3.75 T, $T_{max}$ gets saturated at 2 K due to the limitation in the measurement temperature as the lowest accessible temperature of the system is 2 K. Whereas, the resistance at $T_{max}$ shows a reverse trend with the magnetic field and reaches to the maximum value around 2 k$\Omega$ at 5T at the lowest temperature 2K. Increment in the resistance value at $T_{max}$ with magnetic field is opposite to the phenomenon of WL, where a relatively small magnetic field destroys the constructive quantum interference of scattered electron waves and hence reduction in resistance occurs under magnetic field. Further, the presence of a constant slope in the *R(T)* at high magnetic field and at low temperature is the signature of EEI as evident from Fig. 4(a) & (b). Therefore, it is clear from the field dependent *R(T)* measurements that EEI in the diffusion channel is the main mechanism behind the observed upturn and the associated resistance peak as appeared in the *R(T)*. Moreover, we have carried out *MR* measurements at temperature far above the $T_c$ in order to have an idea about the contribution from WL. However, no trace of negative *MR* which is the signature of WL is observed even at temperature above the *T\** where the superconducting fluctuations can be ignored. We have observed positive *MR* at temperature above the *T\** which confirms the presence of EEI rather than WL. The *MR* at higher temperature for the sample TN10A is shown in Fig. S3 in the SM [24].

Furthermore, from the variation of $T_{max}$ and *T\** with the field, we have constructed a phase diagram which is shown in Fig. 4(d). Here, the extracted temperature points $T_{max}$ & *T\** are observed to shift towards lower temperature under application of a magnetic field and finally, they meet at about 4.75 T which is marked as the crossover field from (QCC) to EEI dominated regime at higher magnetic field.

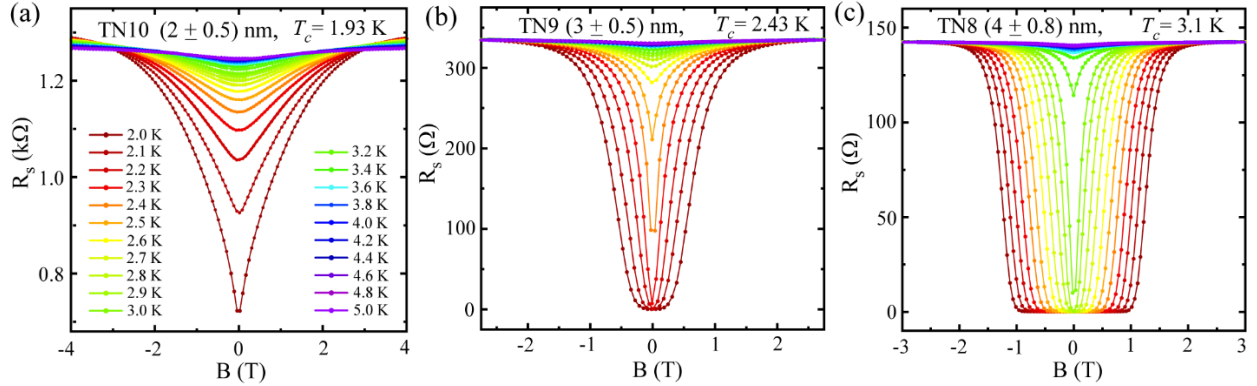

*Fig. 5: Magnetoresistance isotherms for the samples (TN10, TN9 & TN8) annealed at 780 °C and their respective Tc are obtained through total quantum correction to the conductivity (QCC) fitting.*

In addition to the field dependent R(T) measurements, we have carried out isothermal MR measurements to have an insight into the interplay between EEL and the superconducting fluctuations in the presented disordered superconducting TiN thin films. In Fig. 5, we have presented MR isotherms for the samples TN10, TN9 & TN8 annealed at 780 °C and for the rest of the samples (TN4, TN5, TN12 & TN13), the MR isotherms are shown in the SM (Fig. S4) [24]. As mentioned before, the $T_c$ is obtained from the QCC fit using Eq. 6 and all the samples for $T>T_c$ show positive magnetoresistance originated typically from superconducting fluctuations [31]. With increasing temperature, the MR curves shift towards lower magnetic field and the effect of magnetic field gets suppressed. In Fig. 6, we have replotted MR data in terms of magnetoconductance $G(B)=1/R_s(B)$ with $R_s(B)$ as the field dependent sheet resistance. Here, we address mainly three relevant quantum contributions to magnetoconductance (MC), namely, the AL, MT & WL. The Aslamazov-Larkin (AL) contribution comes from the parallel channel formed by the fluctuating cooper pairs [9]. The second contribution Maki-Thompson (MT) is due to the coherent scattering of electrons formed by fluctuating cooper pairs before losing their phase coherence [10, 11]. Here, AL & MT are the two main contributions to conductivity from superconducting fluctuations. AL dominates at temperature close to $T_c$ & MT dominates far above the $T_c$. The third contribution (WL) comes from the constructive quantum interference of scattered electron waves moving in a closed trajectory but opposite in direction [32]. The superconducting fluctuations (AL & MT) give negative &

WL gives a positive contribution to the magnetoconductivity [31]. Now, theoretically, the field-dependent electrical conductance can be expressed as,

$$G_{xx}(B,T) = G^n + \Delta G^{SF}(B,T) + \Delta G^{WL}(B,T) \qquad (7)$$

Where the first term denotes Drude's conductance, the second term represents the conductance from superconducting fluctuations and the last term originates from disorder-induced quantum interference of scattered electronic waves. Here, as quantum contributions are considered, we omit the classical Drude's conductance from Eq. 7 and the total quantum corrections to the magnetoconductivity can be expressed as,

$$\Delta G_{xx}(B,T) = \Delta G^{SF}(B,T) + \Delta G^{WL}(B,T) \qquad (8)$$

First, we address the contributions to magnetoconductance from superconducting fluctuations (AL & MT) and AL contribution to MC can be expressed as,

$$\Delta G^{AL}(B,T) = G_{00} \frac{\pi^2}{8 ln\left(\frac{T}{T_c}\right)} \left\{ 8\left(\frac{B_{SF}}{B}\right)^2 \left[\psi\left(\frac{1}{2} + \frac{B_{SF}}{B}\right) - \psi\left(1 + \frac{B_{SF}}{B}\right) + \frac{B}{2B_{SF}}\right] - 1 \right\} \qquad (9)$$

Where, $\psi(x)$ is the digamma function, $B_{SF}$ characteristic field related to superconducting fluctuations and can be expressed by the Ginzburg-Landau relaxation time $\tau_{GL}$ as, $B_{SF} = \frac{\hbar}{4eD\tau_{GL}}$ with $\tau_{GL} = \frac{\pi\hbar}{8K_B T ln(T/T_c)}$.

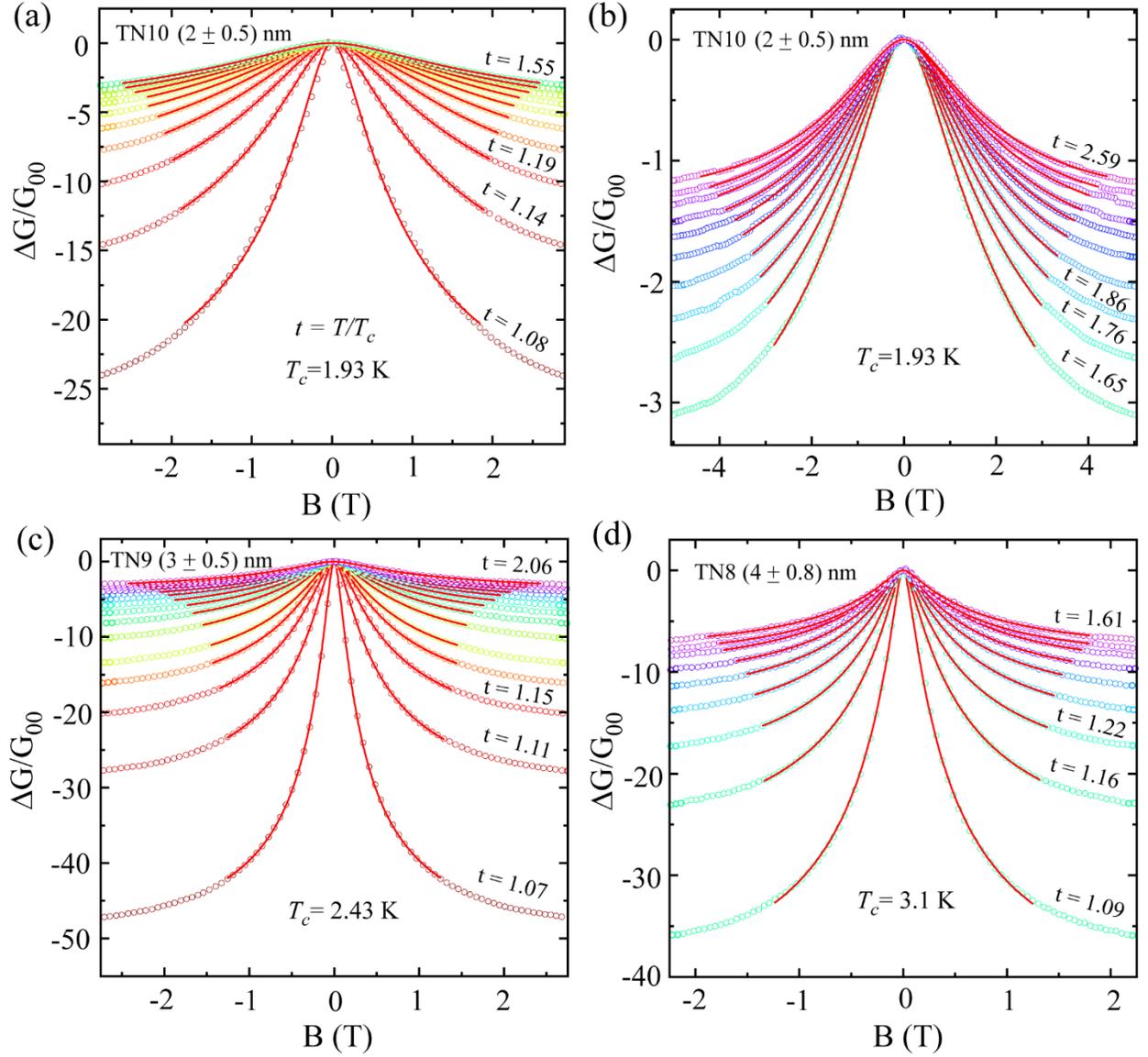

*Fig. 6: Magnetoconductance isotherms for samples TN10 (a & b), TN9(c) & TN8(d). The red solid curves are the fits obtained by the total quantum corrections to the MC that include mainly superconducting fluctuations (AL & MT) & WL as expressed in Eqns. 9, 10 &12, respectively.*

Secondly, MT's contribution to MC can be written in terms of $\beta_{LdS}(T/T_c, \delta)$ which is the magnetic field dependent modified electron-electron attraction strength $\beta(T/T_c)$, the extended version of Larkin's $\beta_L(T/T_c)$ parameter that determines the effective electron-electron interaction strength [16]. $\delta$ is the pair breaking or cut-off parameter [27, 31, 33]. Lopes dos Santos and Abrahams [17] have extended Larkin's results for lower temperatures (close to $ln(T/T_c) \ll 1$)) and higher magnetic field range ($B \ll k_B T/4eD$) [17, 31, 33] because Larkin's calculation on the temperature and magnetic field exclude the immediate vicinity of $T_c$ and is subjected to low magnetic field. Therefore, the extended form of *MT* contribution from Lopes dos Santos and Abrahams can be written as [17, 33].

$$\Delta G^{MT}(B,T) = -G_1 \cdot \frac{e^2}{2\pi^2 \hbar} \left[ \psi\left(\frac{1}{2} + \frac{B_\phi}{B}\right) - \psi\left(\frac{1}{2} + \frac{B_{SF}}{B}\right) + \ln\left(\frac{B_{SF}}{B_\phi}\right) \right] \quad (10)$$

Where, $G_1 = \beta_{LdS}(T/T_c, \delta)$, $\beta_{LdS}(T/T_c, \delta) \equiv \pi^2/4(\epsilon - \delta)$, $\epsilon \equiv \ln\left(\frac{T}{T_c}\right)$, $\delta = \frac{\pi \hbar}{8 k_B T} \frac{1}{\tau_\phi}$ (11)

Thirdly, WL's contribution to MC can be written as,

$$\Delta G^{WL}(B,T) = N \cdot \frac{e^2}{2\pi^2 \hbar} \cdot Y\left(\frac{B}{B_\phi}\right), \quad (12)$$

Where, $Y(x) = \ln(x) + \psi\left(\frac{1}{2} + \frac{1}{x}\right)$ and $B_\phi = \frac{\hbar}{4 e D \tau_\phi}$, $D = \frac{\pi}{2\gamma} \frac{k_B T_c}{e B_{c2}(0)}$, $\gamma = 1.78$, (13)

Here, $\tau_\phi$ is dephasing scattering time and the associated magnetic field ($B_\phi$) is known as the phase-breaking field, and *D* is the diffusion constant. The coefficient *N* in equation (12) represents the number of channels participating in the conduction process [28]. Further, we have calculated magnetoconductance from the measured longitudinal magnetoresistance with resistance in Ohm/Square by using the given expression,

$$\Delta G_{xx}(B) = \frac{1}{R_{xx}(B)} - \frac{1}{R_{xx}(0)} = -\frac{R_{xx}(B) - R_{xx}(0)}{R_{xx}(B) \cdot R_{xx}(0)} \quad (14)$$

and have plotted them in unit of $G_{00} = \frac{e^2}{2\pi^2 \hbar}$ as, $\Delta G_{xx}(B)/G_{00}$ in Fig. 6.

Here, we have taken care of the dimensionality as discussed above & validity of MT expression ($B \ll k_B T/4eD$) in selecting the magnetic field range for fitting the experimental data shown in Fig. 6. Combining the AL, MT & WL contributions by using Eqns. 9, 10 & 12, respectively, we have fitted the experimental magnetoconductance for temperature above $T_c$, i.e., $t > 1$, where $t = T/T_c$, is the reduced temperature. For the fits as shown by the red solid curves in Fig. 6, the characteristic fields $B_\phi$ and $B_{SF}$ have been used as the free parameters and $\beta_{LdS}(T/T_c, \delta)$ is exactly taken as it is expressed in Eq. 11. The fits show excellent agreement with the experimental data as shown in Fig. 6. For a clearer view, the magnetoconductance isotherms for the sample TN10 are split into two sets based on the temperature range. For the set with $t$ in the range from 1.08 to 1.55 is shown in Fig. 6(a) and the second set up to $t$ =2.59 is shown in Fig. 6(b). For TN9 and TN8, the MC data along with the fits are presented in Fig 6(c) and (d), respectively. The MC curves and the corresponding fits for rest of the samples belonging to $T_a$ (820°C & 750°C) are shown in the SM (Fig. S5) [24]. From the fit, $\tau_\phi$ has been extracted through the phase breaking field $B_\phi$. Further, the coefficient for the MT contribution to the MC, $\beta_{LdS}(T/T_c, \delta)$, as expressed in Eq. 10 & 11, relates with $\tau_\phi$ via the pair breaking parameter $\delta$. For all the samples, we have evaluated the inverse phase scattering time $(\tau_\phi^{-1})$ and $\beta_{LdS}(T/T_c, \delta)$ and they have been plotted with respect to the reduced temperature $T/T_c$ in Fig. 7. Here, $\beta_{LdS}(T/T_c, \delta)$ is the parameter introduced by Lopes dos Santos and Abrahams in the modified version of Larkin's theory [17]. The extended version of the theory is valid for the temperature points close to the $T_c$, where ($ln(T/T_c) \ll 1$). In our case, all the temperature points are adhered to the condition mentioned for the validity of extended version of Larkin's theory [17]. In the limit, $(ln(T/T_c) \ll 1)$, the $\beta_{LdS}(T/T_c, \delta)$ takes the form as mentioned in Eq. 11 and depends on $\tau_\phi$ values. Interestingly, the variation of $\beta_{LdS}(T/T_c, \delta)$ on the reduced temperature, as observed in Fig. 7(a), merge on top of each other for all the measured samples presented in this work. Hence, the dependence of $\beta_{LdS}(T/T_c, \delta)$ on the reduced temperature follows a unanimous trend and doesn't depend on the growth parameters as long as the two-dimensionality is maintained. In this case, the growth parameters are mainly the annealing temperature ($T_a$) and the sample thickness ($d$). As evident in Fig. 7(a), the variation of $\beta_{LdS}(T/T_c, \delta)$ shows weak dependence or becomes almost independent of temperature for $T/T_c > 1.5$ and it diverges as $T$ approaches to $T_c$. In order to have the confirmation of this generalized trend in the dependence of $\beta_{LdS}(T/T_c, \delta)$ on the reduced temperature, inset of Fig. 7(a) magnifies the diverging part of $\beta_{LdS}(T/T_c, \delta)$ close to $T_c$ and indeed, all the samples follow the common path in the plot.

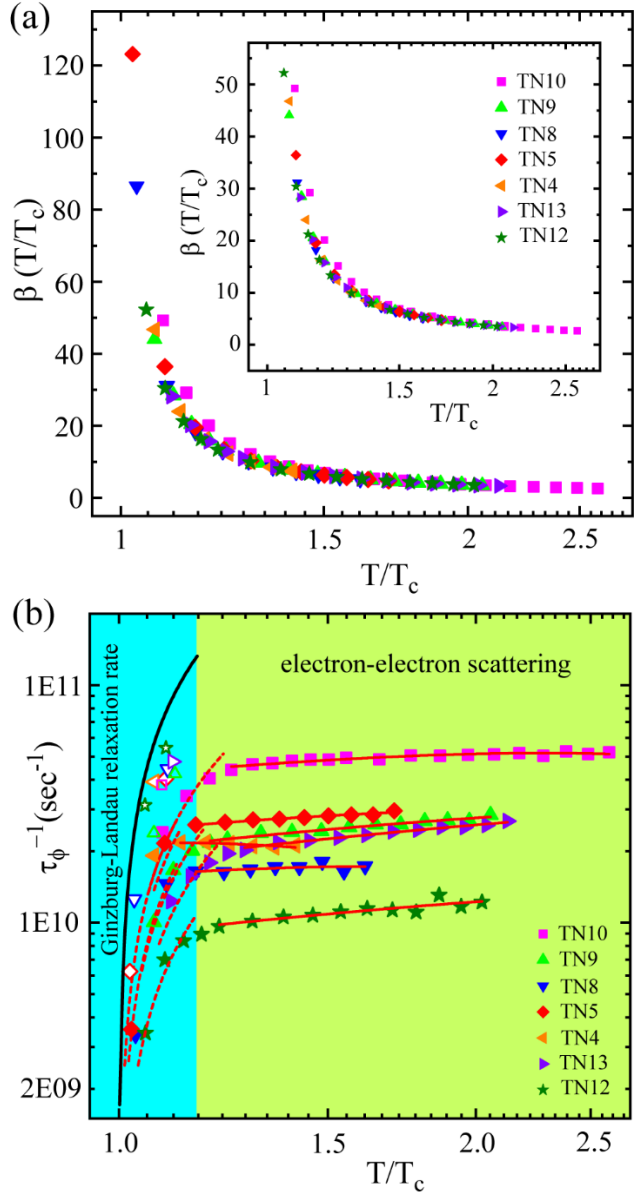

*Fig. 7:* Dependence of modified Larkin's parameter $\beta_{LdS}(T/T_c, \delta)$ and inverse phase scattering time $(\tau_\phi^{-1})$ on reduced temperature $T/T_c$. (a) A semi-logarithmic presentation of the variation of $\beta_{LdS}(T/T_c, \delta)$ with reduced temperature $(T/T_c)$ for all the TiN samples presented in this work from different batches annealed at 820 °C, 780 °C & 750 °C and with different film thickness. The inset shows the magnified view of $\beta_{LdS}(T/T_c, \delta)$ variation with the reduced temperature for T close to $T_c$. (b) Logarithmic plot of inverse phase scattering time $(\tau_\phi^{-1})$ with reduced temperature for all the measured TiN samples. The red dashed curves correspond to $\tau_\phi^{-1} \propto (T - T_c)$, the black solid curve represents the Ginzburg-Landau relaxation rate $\tau_{GL}^{-1} = (8k_B/\pi\hbar)(T - T_c)$ with $T_c$=2.43 K and the solid red curves are the fits using Eq.15 for inverse electron-electron scattering time. The open symbols represent the inverse characteristic time obtained from the characteristic field $B_{SF}$.

Furthermore, the extracted values of $B_\phi$ from the magnetoconductance fitting are converted into $\tau_\phi$ by using the expression $B_\phi = \frac{\hbar}{4eD\tau_\phi}$ for all the TiN samples and they are presented as inverse phase scattering time ($\tau_\phi^{-1}$) with respect to the reduced temperature in Fig. 7(b). Here, also we observe that the samples follow a general trend with two distinct regions with clear variation of the slopes that are highlighted by two different shades. The first region, as highlighted by the cyan shade (close to $T_c$), shows an abrupt decrement in $\tau_\phi^{-1}$ as temperature approaches to $T_c$ from high temperature. The second region highlighted in light green shade for $T/T_c > 1.17$ demonstrates a linear variation of $\tau_\phi^{-1}$ with temperature. Generally, there are three inelastic scattering mechanisms that lead to the phase relaxation in 2D superconductors in dirty limit and the corresponding scattering rates are (i) the electron-phonon scattering rate $\tau_{e-ph}^{-1}$ (ii) the inelastic electron-electron scattering rate $\tau_{e-e}^{-1}$ and (iii) the inelastic scattering rate of electrons due to superconducting fluctuations $\tau_{e-fl}^{-1}$ [34]. The electron-phonon scattering rate ($\tau_{e-ph}^{-1} \propto T^3$) is mainly prominent at high temperature [27, 35]. As the magnetotransport study presented in the work has been carried out for the temperature range $T \leq 2.5T_c$ where superconducting fluctuations and electron-electron interaction mechanisms are dominant, the electron-phonon scattering can be ignored [13]. Further, for a dirty superconductor in 2D limit where the thermal diffusion length is more than the sample thickness (which is the case for the present set of samples as listed in Table-1), the inelastic electron-electron scattering rate $\tau_{e-e}^{-1}$ varies linearly with temperature according to the expression [13, 34],

$$\tau_{e-e}^{-1} = \frac{e^2 R}{2\pi\hbar^2} k_B T \ln \frac{\pi\hbar}{e^2 R} \qquad (15)$$

The experimental data points show excellent agreement with the theoretical expression given in Eq. 15 for inelastic electron-electron scattering as shown by the red solid curves in Fig. 7(b) for the light green shaded region. However, the region below $1.17T_c$

As the temperature approaches to $T_c$, the inelastic scattering of electrons due to superconducting fluctuations becomes prominent and the electron-fluctuation rate $\tau_{e-fl}^{-1}$ is given by [36, 37],

$$\tau_{e-fl}^{-1} = \frac{e^2 R_S}{2\pi\hbar^2} k_B T \frac{2ln2}{ln\frac{T}{T_c}+D}, \text{ with}$$

$$D = \frac{4ln2}{\sqrt{ln^2\left(\frac{\pi\hbar}{e^2 R_S}\right)+\frac{128\hbar}{e^2 R}} - ln\left(\frac{\pi\hbar}{e^2 R_S}\right)} . \quad (16)$$

According to Eq. 16 a strong upturn in the temperature dependence of $\tau_\phi^{-1}$ at $T$ very close to $T_c$ is predicted as $ln(T/T_c)$ in the denominator of the $\tau_{e-fl}^{-1}$ term becomes almost *zero*. However, here in the present study, we observe a strong downturn in $\tau_\phi^{-1}$ as the temperature approaches to $T_c$ from higher temperature. Similar behavior of $\tau_\phi^{-1}$ near $T_c$ has been also observed for In/InO$_x$ composite films [38], Re$_{70}$W$_{30}$ and Nb$_{1-x}$Ta$_x$ thin films [39, 40]. The phase relaxation rate $\tau_\phi^{-1}$ in this regime varies proportionally to $(T - T_c)$ as shown by the red dashed curves in Fig. 7(b) in a similar fashion as that of the Ginzburg-Landau (GL) phase relaxation rate $\tau_{GL}^{-1} = (8k_B/\pi\hbar)(T - T_c)$ at $T$ very close to $T_c$ [40].

In the immediate vicinity of $T_c$, the resemblance in temperature variation of $\tau_\phi^{-1}$ with that of $\tau_{GL}^{-1}$ can be understood from the AL contribution which is one of the most prominent contributions to the MC near $T_c$ [41]. The AL term is fundamentally different from the other terms as the only characteristic field $B_{SF}$ (Eq. 9) used in AL term is different than the characteristic dephasing field $B_\phi$ used in other terms [41]. The characteristic field $B_{SF}$ is associated with the Ginzburg-Landau time $\tau_{GL} = \frac{\pi\hbar}{8k_B T ln(T/T_c)}$ [40]. Hence the phase relaxation rate $\tau_\phi^{-1}$ obtained from the AL term near $T_c$ can measure the GL phase relaxation rate $\tau_{GL}^{-1}$ [40]. In Fig. 7(b), the solid black curve for $T/T_c <1.17$ represents the $\tau_{GL}^{-1}$ for $T_c = 2.43$ K and the open symbols represent the inverse characteristic time related to $B_{SF}$. It is clear from the figure that the phase relaxation rate related to $B_{SF}$ closely follow the GL phase relaxation rate $\tau_{GL}^{-1}$ for the set of samples presented in Table-1. The deviation for the experimental points from the GL rate $\tau_{GL}^{-1}$ as appeared in Fig. 7(b) can be originating from some contributions from normal electrons that are modified due to superconducting fluctuations near the transition [42]. Therefore, clearly the first regime with steeper slope relates to dephasing caused by superconducting fluctuations whereas, the second regime corresponds to inelastic electron-electron scattering induced dephasing and all the samples offer the generic trend as evident in Fig. 7(b).

## IV. CONCLUSION

To summarize, we have revisited the quantum corrections to the conductivity (QCC) terms for disordered 2D superconducting TiN thin films. We observe a strong interplay between the superconducting fluctuations and electron-electron interactions. The *R(T)* measurements, carried out under zero magnetic field, feature different regimes that are mainly defined by the sign of the slope *dR/dT*. Transitions from metallic to weak insulating and further to superconducting regime are obtained for the samples while cooling down from room temperature to 2 K. In spite of having different growth conditions such as annealing temperature and film thickness, all the samples presented in this article follow a similar trend in their zero-field *R(T)* characteristics that feature a resistance dip at the end of the metallic state, an upturn along with a resistance peak indicating a weak insulating state and finally a sharp drop in the resistance due to the onset of superconductivity. We have shown that the intermediate upturn and hence the weak-insulating type of behavior in the zero-field *R(T)* is mainly due to the electron-electron interaction which is further supported by the field dependent *R(T)*. However, the samples presented here are of low resistance (< 1,5 kΩ) values and hence, the observed upturns are also weak. In order to observe a stronger upturn, we have measured a highly resistive sample with $R_{max}$ ~ 18 kΩ which eventually offers a much stronger upturn in the zero-field *R(T)*. The results from the zero-field *R(T)* and *R(T)* under external magnetic field (Fig. S6 in the supplemental Material) [24] are consistent with that obtained from the low resistive samples presented here.

Here, weak localization does not play any significant role as supported by the MR measurements also. Further, from the analysis of AL, MT & WL contributions to the magnetoconductance, we have investigated the variations of the modified Larkin parameter *β(T/T$_c$)* and the inverse phase relaxation time ($\tau_\phi^{-1}$) on the reduced temperature (*T/T$_c$*) and both of them show a generic pattern which all the samples follow very closely.

## ACKNOWLEDGEMENTS

We highly acknowledge UGC-DAE CSR, Indore, India for carrying out the low temperature resistivity measurements in PPMS. We are thankful to Dr. Sudhir Husale for his critical reading of the manuscript and also for his invaluable suggestions on the manuscript. S.Y. acknowledges the Senior Research fellowship (SRF) from UGC. This work was supported by CSIR network project 'AQuaRIUS' (Project No. PSC 0110) and is carried out under the mission mode project "Quantum Current Metrology"

*Supplemental Material*

# Probing electron-electron interaction along with superconducting fluctuations in disordered TiN thin films

*Sachin Yadav,* [1,2] *Vinay Kaushik,* [3] *M.P. Saravanan,* [3] *and Sangeeta Sahoo\**[1, 2]

[1]*CSIR-National Physical Laboratory, Dr. K. S. Krishnan Marg, New Delhi-110012, India*

[2]*Academy of Scientific and Innovative Research (AcSIR), Ghaziabad- 201002, India*

[3]*Low Temperature Laboratory, UGC-DAE Consortium for Scientific Research, University Campus, Khandwa Road, Indore 452001, India*

**Contents:**

1. **Zero-field R(T) data for samples TN5, TN4, TN13, TN12**
2. **Logarithmic temperature dependence of the zero-field conductivity for samples TN5, TN4, TN13, TN12**
3. **Magnetoresistance data measured at high temperature for a sample TN10A (taken from the batch of TN10).**
4. **Field dependent resistance R(B) isotherms for samples TN5, TN4, TN13, TN12:**
5. **Fitting analysis of magnetoconductance data by using all the quantum contributions for samples TN5, TN4, TN13, TN12:**
6. *Table 1: The fitting parameters obtained from the total quantum corrections to the conductivity (QCC) fits to the zero-field R(T) as shown in Fig. 2 & Fig. S1.*
7. *The R(T) measured under zero-field and also under external field for a higher resistive sample.*
8. **Atomic force microscopy (AFM) images for the characterization of the surface morphology of the samples presented in the main manuscript.**

➢ **Zero-field R(T) data for samples TN5, TN4, TN13, TN12:**

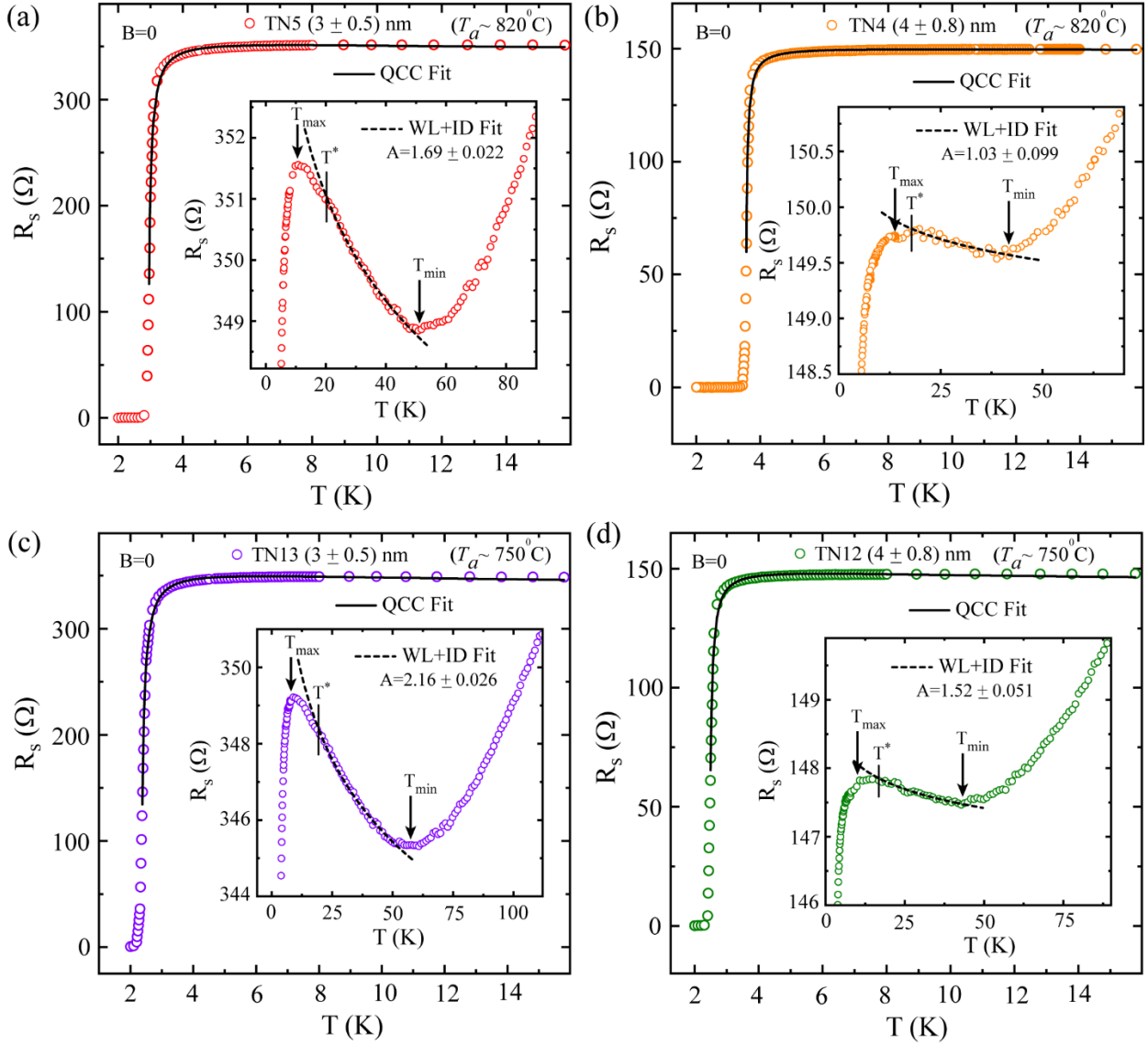

*Fig. S1: A set of zero field temperature dependent resistance [R(T)] data of TiN thin films annealed at 820°C & 750°C with different film thickness fitted with total QCC theory. (a-d) The (RT) data fitted with black solid lines with total QCC theory and inset demonstrates the resistive upturn in (RT) from $T_{min}$ to $T_{max}$ and fitted with (WL+ID) model shown by black dashed lines. $T_{min}$, $T_{max}$ & $T^*$ are shown in the inset of Fig. (a-d), where $T_{max}$ & $T_{min}$ are the maximum and minimum temperature points of the upturn curvature and $T^*$ is the point from where (WL+ID) fitting starts to divert from its normal behavior. Here, A is the fitting parameter derived from the (WL+ID) theoretical fitting.*

- **Logarithmic temperature dependence of the zero-field conductivity for samples TN5, TN4, TN13, TN12:**

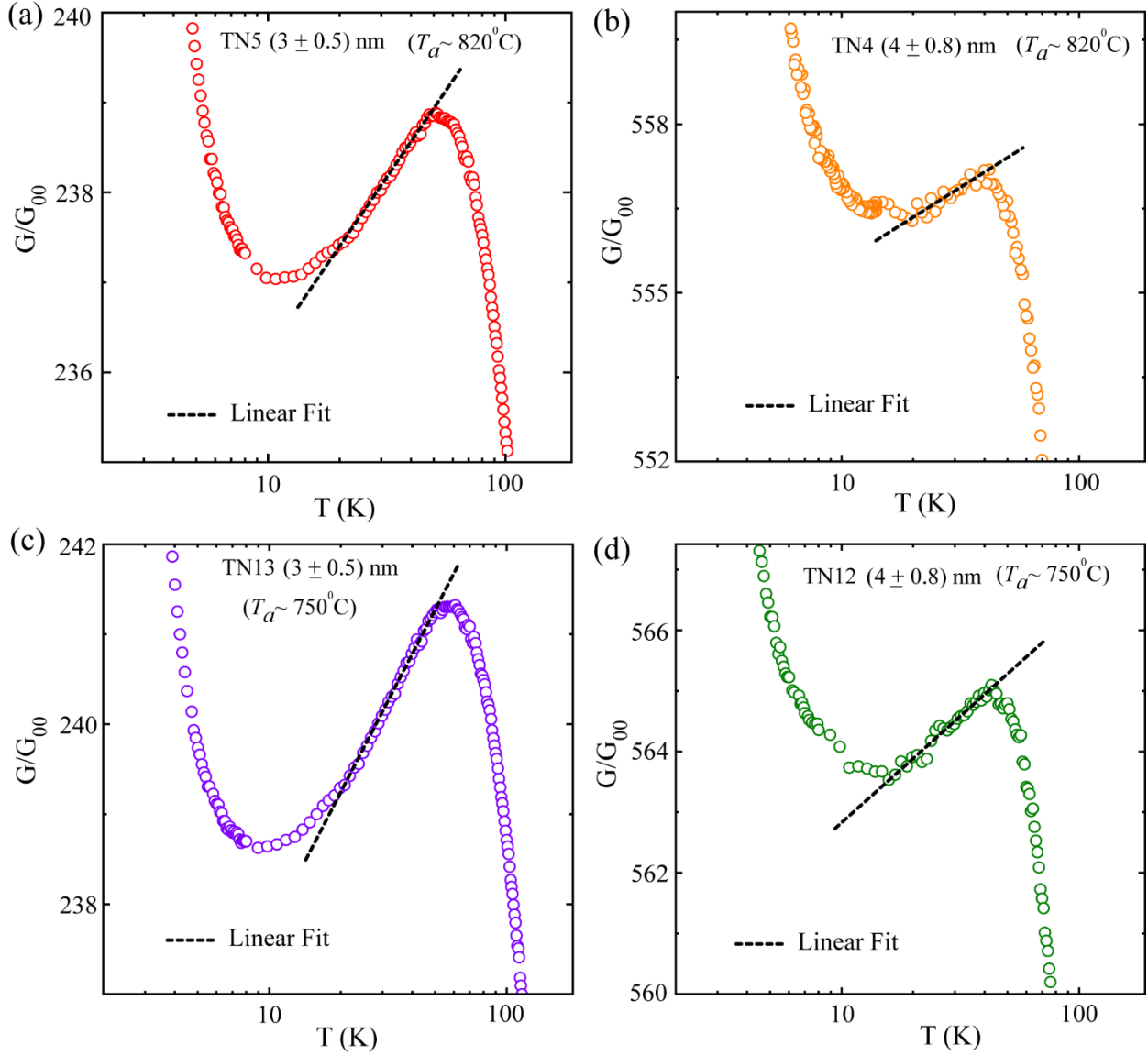

*Fig. S2: Logarithmic temperature dependence of dimensionless conductance in semi-log scale for the samples TN4, TN5, TN12 & TN13 annealed at 820°C & 750°C with different film thickness and black solid dashed lines are linear fits to the experimental data.*

➤ **Magnetoresistance data measured at high temperature for a sample TN10A (taken from the batch of TN10).**

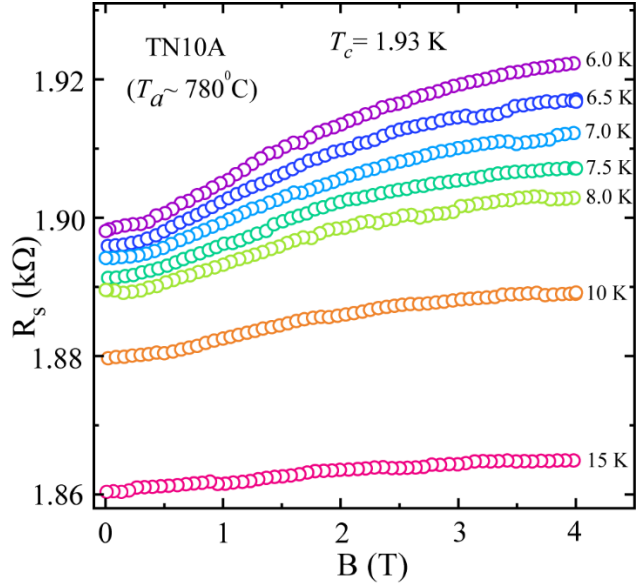

*Fig. S3: Magnetoresistance (MR) isotherms for the sample TN10A with film thickness around 2 nm showing positive MR at high temperature far away from $T_c$ and even above $T^*$.*

Magnetoresistance (*MR*) isotherms demonstrate the positive MR at high temperature where the contributions of superconducting fluctuations become minimum and the corresponding temperature is marked as $T^*$ in the main manuscript. Above $T^*$ the electronic transport is dominated by weak localization (WL) and electron-electron interaction (EEI) in diffusion channel (ID). Moreover, the appearance of negative *MR* is the hallmark of weak localization above $T^*$, where contribution from the superconducting fluctuation can be totally ignored. However, it is evident from Fig. S3 that there is no trace of negative *MR* above $T^*$ and appearance of positive MR indicates towards the contribution of EEI in the resistive upturn.

➢ **Field dependent resistance R(B) isotherms for samples TN5, TN4, TN13, TN12:**

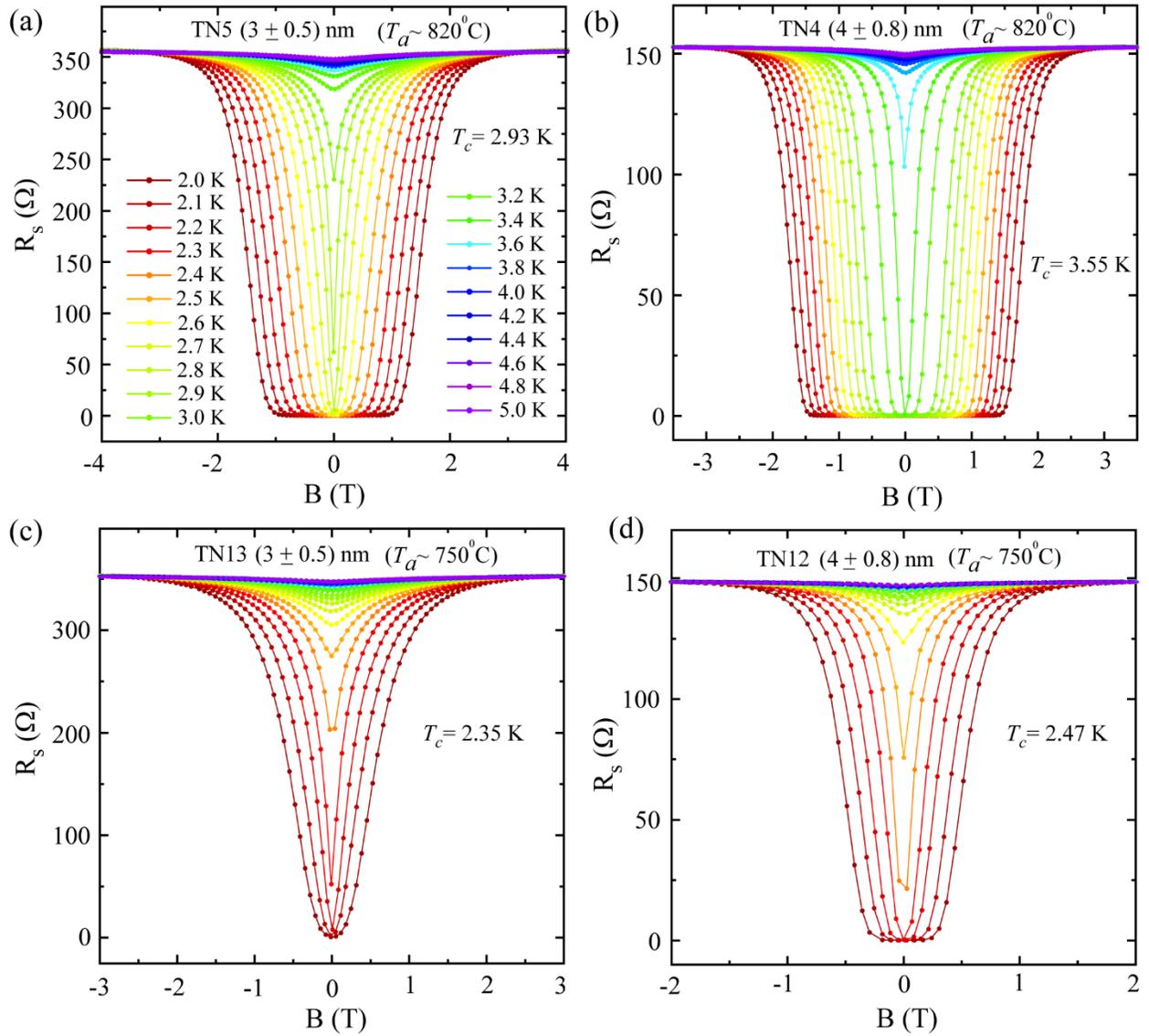

*Fig. S4: MR isotherms for the samples TN4 & TN5 annealed at 820°C and for TN12 & TN13 annealed at 750°C with different film thickness. Their respective $T_c$ values are obtained through total quantum correction to the conductivity (QCC) fitting.*

- **Fitting analysis of magnetoconductance data by using all the quantum contributions for samples TN5, TN4, TN13, TN12:**

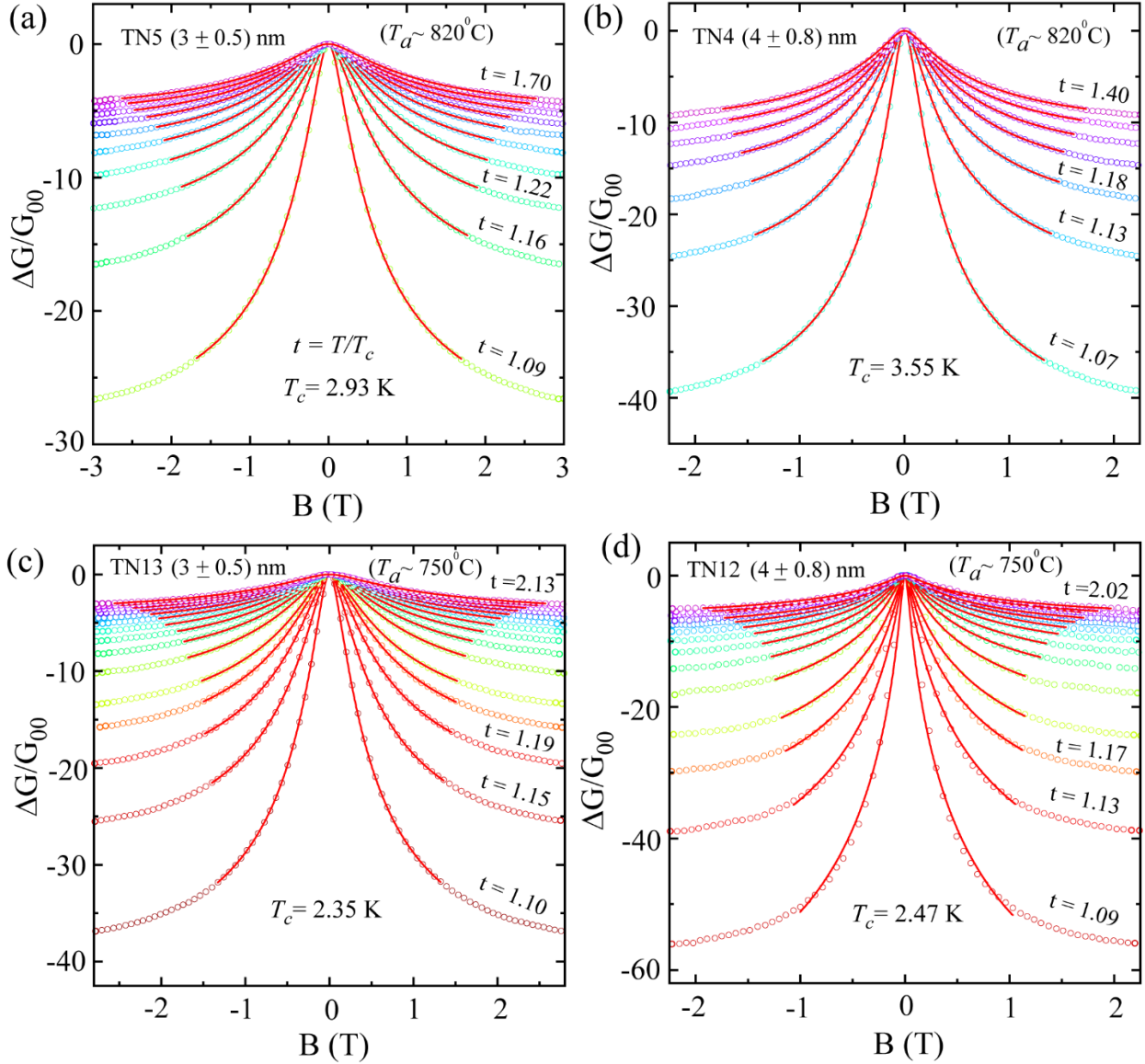

*Fig. S5: A set of isothermal magnetoconductance for the samples TN4 & TN5 annealed at 820°C and for TN12 & TN13 annealed at 750°C with different film thickness. The red solid curves are the fits obtained through the combined contributions from superconducting fluctuations (AL&MT) & WL by using the Eq. 9,10 &12.*

➢ *Table 1: The fitting parameters obtained from the total quantum corrections to the conductivity (QCC) fits to the zero-field R(T) as shown in Fig. 2 & Fig. S1.*

| Samples | A | $\tau$ (elastic scattering time), (fs) | $\tau_\phi$ (inelastic scattering time), (ps) |
|---|---|---|---|
| TN4 | 1.03 ± 0.099 | 9.95 ± 3.16 | 7.70 ± 0.14 |
| TN5 | 1.69 ± 0.022 | 5.17 ± 0.26 | 7.48 ± 0.16 |
| TN8 | 0.75 ± 0.051 | 10.5 ± 2.99 | 13.6 ± 0.39 |
| TN9 | 2.21 ± 0.018 | 7.58 ± 0.21 | 14.6 ± 0.39 |
| TN10 | 1.70 ± 0.01 | 8.59 ± 0.30 | 7.23 ± 0.21 |
| TN12 | 1.52 ± 0.051 | 6.41 ± 0.82 | 61.7 ± 4.55 |
| TN13 | 2.16 ± 0.026 | 7.12 ± 0.31 | 18.5 ± 0.59 |

- *The R(T) measured under zero-field and also under external field for a higher resistive sample.*

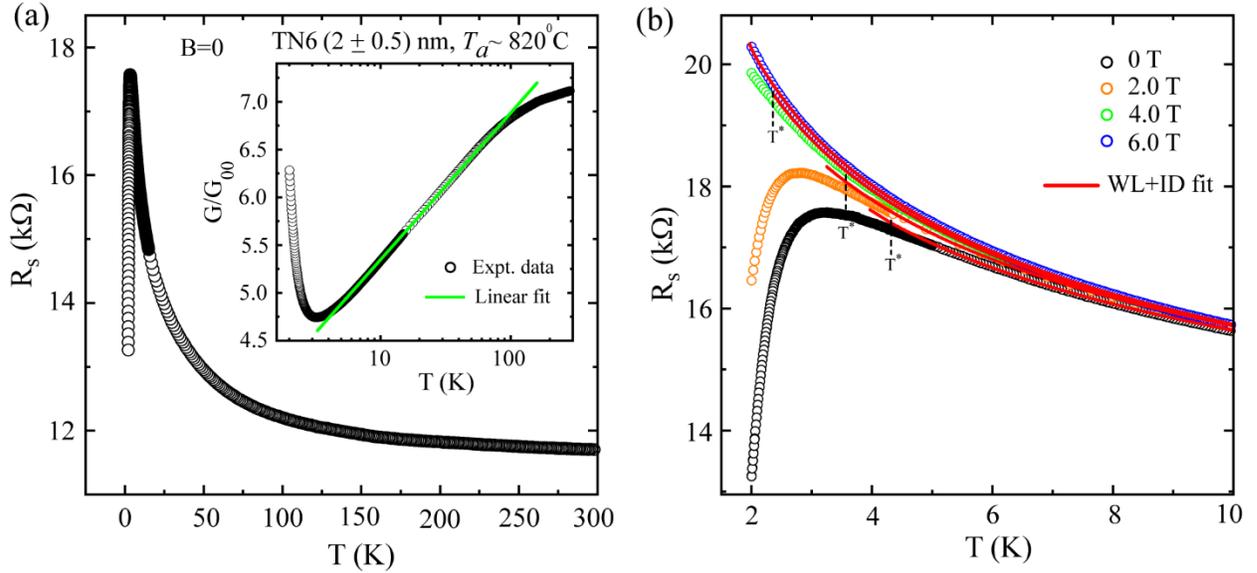

*Fig. S6: R(T) measurements for a highly resistive sample TN6. (a) Zero-field R(T) data showing a strong upturn in resistance. Inset: The logarithmic dependence of the zero-field conductance. (b) R(T) data measured under magnetic field applied perpendicular to the sample plane. The dashed vertical lines represent the characteristic temperature T* above which EEI dominates. The red curves correspond to WL+ID fits.*

➢ **Surface morphology by atomic force microscopy (AFM) images for the samples presented in this work**

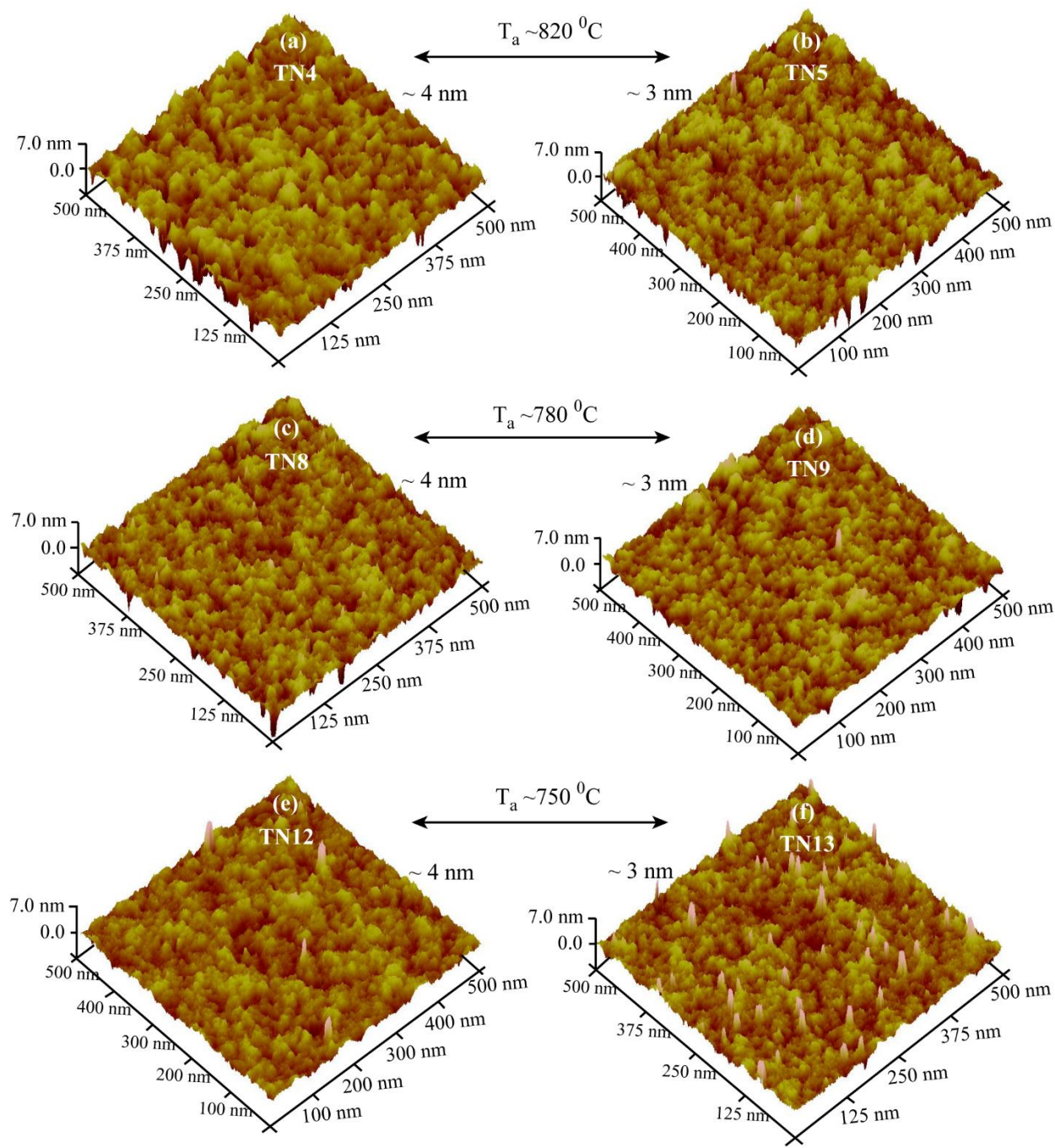

*Fig. S7*: *Atomic force microscopy (AFM) images for the characterization of the surface morphology of the samples presented in the main manuscript.*